# Molecular DFT+U: A Transferable, Low-Cost Approach to Eliminate Delocalization Error

Akash Bajaj[1,2] and Heather J. Kulik[1,*]

[1]Department of Chemical Engineering, Massachusetts Institute of Technology, Cambridge, MA, 02139

[2]Department of Materials Science and Engineering, Massachusetts Institute of Technology, Cambridge, MA, 02139

AUTHOR INFORMATION

**Corresponding Author**

*email: hjkulik@mit.edu, phone: 617-253-4584

ABSTRACT While density functional theory (DFT) is widely applied for its combination of cost and accuracy, corrections (e.g., DFT+U) that improve it are often needed to tackle correlated transition-metal chemistry. In principle, the functional form of DFT+U, consisting of a set of localized atomic orbitals (AO) and a quadratic energy penalty for deviation from integer occupations of those AOs, enables the recovery of the exact conditions of piecewise linearity and the derivative discontinuity. Nevertheless, for practical transition-metal complexes, where both atomic states and ligand orbitals participate in bonding, standard DFT+U can fail to eliminate delocalization error (DE). Here, we show that by introducing an alternative valence-state (i.e., molecular orbital or MO) basis to the DFT+U approach, we recover exact conditions in cases where standard DFT+U corrections have no error-reducing effect. This MO-based DFT+U also eliminates DE where standard AO-based DFT+U is already successful. We demonstrate the transferability of our approach on a range of ligand field strengths (i.e., from $H_2O$ to CO), electron configurations (i.e., from Sc to Fe to Zn), and spin states (i.e., low-spin and high-spin) in representative transition-metal complexes.

**TOC GRAPHICS**

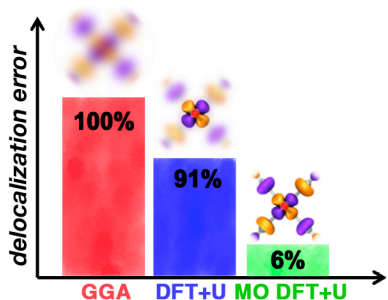

Applied Kohn–Sham density functional theory (DFT)[1-4] is the electronic structure method of choice across chemistry and materials science due to its good balance of cost and accuracy[5-6]. However, widely used local[7-8] and semi-local[9-11] approximations to the exchange-correlation (xc) functional within DFT are plagued by one- and many-electron self-interaction errors (SIEs).[12-16] The many-electron SIE can be tied back to the fact that these approximations lack key properties of the exact functional, including a derivative discontinuity and piecewise linearity upon fractional charge addition[17-23]. This has often been interpreted as a delocalization error (DE)[24-27] that can be tied to erroneous predictions of densities[24,28-32], electron affinities[33-36], band gaps[27,37-38], spin-state ordering[39-47] and other properties[48-51]. Among the possible generalizations[52-53] to Kohn–Sham DFT that help mitigate these errors[51,54-66], the DFT+U approach[67-71] is commonly employed for transition-metal-containing systems[39,72-83] as it allows a targeted accounting of electronic correlation of the localized electrons (i.e., $d$ or $f$) at the cost of semi-local DFT[70].

The simplified[69,71], rotationally invariant[68] form of DFT+$U$ is:

$$E^{\text{DFT+U}} = E^{\text{DFT}} + \frac{1}{2} \sum_{I,\sigma} \sum_{nl} U_{nl}^{I} \left[ \text{Tr}\left( \mathbf{n}_{nl}^{I\sigma} \left(1 - \mathbf{n}_{nl}^{I\sigma}\right) \right) \right] . \qquad (1)$$

where a +U correction is applied on every $nl$ subshell and spin $\sigma$ on atom $I$. The occupation matrix $\mathbf{n}_{nl}^{I\sigma}$ is obtained by projecting states $|\psi_{k,v}\rangle$ onto localized atomic orbitals (AOs) from the $nl$ subshell with angular momenta $m$ $|\phi_m^I\rangle$ on site $I$ using:

$$n_{mm'}^{I\sigma} = \sum_{k,v} \langle \psi_{k,v} | \phi_{m'}^I \rangle \langle \phi_m^I | \psi_{k,v} \rangle . \qquad (2)$$

The deviation from linearity of the energy[17] (i.e., the energetic delocalization error, EDE[28]) is approximately quadratic for most exchange-correlation (xc) functionals[18,84]. Thus, the first-principles motivation for using DFT+U to correct semi-local DFT errors is made clear by the fact that the quadratic DFT+U expression in eq. (1) should recover piecewise linearity[71]. The only caveat to this analysis is that we require the fractional electron addition to be captured by the projection onto AOs in $\mathbf{n}_{nl}^{I\sigma}$.[84] For the simplifying case where electron addition produces quadratic energy dependence and is localized to a single element of the occupation matrix, the $U$ value that recovers piecewise linearity[84-86] is the constant energetic curvature[84] of the original xc functional. The curvature[18,84] value to apply as the $U$ value in DFT+U can be obtained from first-principles[87-88] as the difference between the eigenvalues of the highest occupied molecular orbital (HOMO) of the $N+1$-electron system and the lowest unoccupied molecular orbital (LUMO) from the $N$-electron system with the original xc functional:

$$\left\langle \frac{\partial^2 E}{\partial q^2} \right\rangle = \varepsilon_{N+1}^{\text{HOMO}} - \varepsilon_{N}^{\text{LUMO}} \quad . \tag{3}$$

Even if deviations from linearity exhibit higher-order dependence on fractional electron number, this average curvature often remains a good approximation.[18] However, for many cases even with relatively ideal behavior (i.e., quadratic EDE and a single atomic orbital for electron addition), the practical $U$ value needed to recover linearity is higher[84] than the curvature value. Further, standard DFT+U can fail completely and have no effect on EDE for pathological cases[84,89]. Three types of error drive failure of standard DFT+U, depending on the orbitals to which fractional charge is added: i) multiple AO-like MOs, ii) MOs that rehybridize with electron addition, or iii) strongly hybridized MOs that do not correspond to well-localized 3$d$ AOs. Given the reliance of many high-throughput screening workflows on DFT+U to improve properties without increasing cost[90], a robust low-cost alternative is needed. In this Letter, we develop an approach to improve standard DFT+U to ensure the elimination of EDE without losing its favorable scaling or transparent influence on frontier orbital energies and occupations. We achieve this by introducing the use of alternative molecular orbital basis for projectors that adapt to the presence of hybridization.

For typical mononuclear octahedral transition-metal complexes, we previously noted several pathological cases where $U$ values close to those derived from the expression in eq. (3) had no effect at all on reducing semi-local DFT EDE[84]. A representative pathological case[84] is the $[Mn(CO)_6]^{(3-q)+}$ complex in its high-spin (HS) state along the path from Mn(III) to Mn(II) (i.e., $q = 0$ to 1). Significant type-iii error due to metal–ligand hybridization makes the standard DFT+U approach with AOs ineffective in eliminating delocalization error (Figure 1). Quantitatively, the HOMO of the reduced $[Mn(CO)_6]^{2+}$ complex consists of comparable contributions from the $Mn(3d_{x^2-y^2})$ AO (51%) and the coordinating C atom $s$ and $p$ valence AOs (39%, Figure 1 and Table S1).

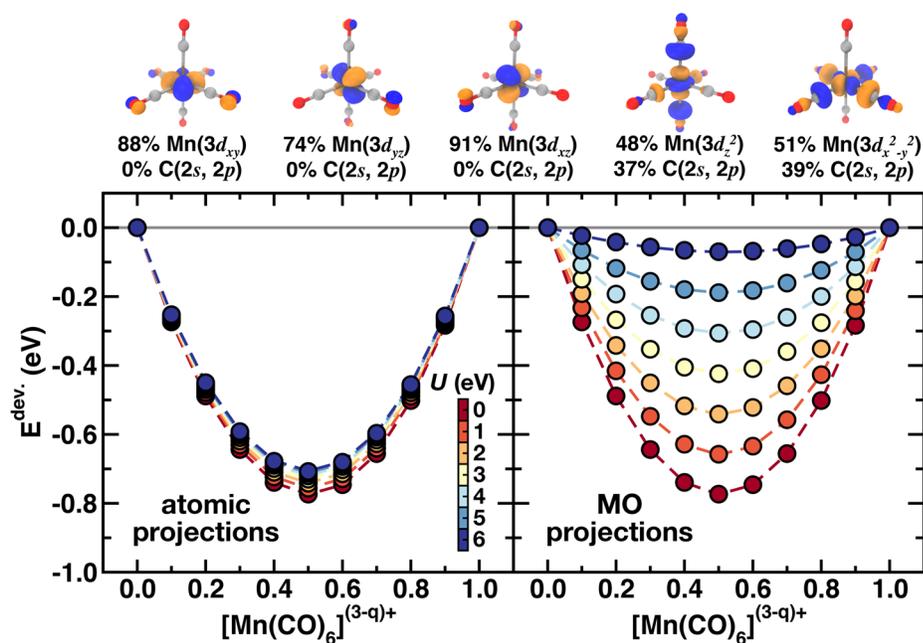

**Figure 1.** Deviations from linearity ($E^{dev.}$, in eV) for HS $[Mn(CO)_6]^{(3-q)+}$ from $q = 0$ to 1 using DFT+U with AO (left) and MO (right) projections. Both $E^{dev.}$ values obtained from fractional charge calculations (filled circles) and spline interpolation (dashed lines) are shown from $U = 0$ to 6 eV in integer increments, as in inset colorbar. MO projectors were obtained from the reduced (i.e., $q = 1$) complex. Density isosurfaces ($|0.002|$ e/bohr$^3$, positive phase in blue and negative phase in orange) of the five spin-up projector MOs are shown at top with the contributions from Mn $3d$ or C valence (i.e., $2s$ or $2p$) annotated.

To overcome the limits of conventional DFT+U with AO projections, we compute alternative projectors from real-space representations of frontier MOs of HS $[Mn(CO)_6]^{2+}$ (Figure 1 and Text S1 and see Computational Details). Since we must choose a single oxidation state for obtaining the projectors, we select the reduced, $N+1$-electron (i.e., $q = 1$) endpoint to favor using occupied orbitals for the MO projectors. Formally, the choice of projector oxidation state is also inherent to the AOs widely employed in standard DFT+U, because they are derived from the all-electron calculation of an isolated ion in a fixed-charge state obtained during pseudopotential generation and are known to have oxidation state dependence.[91-92] To choose which MO states to include, we note the fractional charge line varies between a formally sextet $d^5$ Mn(II) to quintet $d^4$ Mn(III), and so we select the five spin up frontier states (i.e., HOMO-4 to HOMO) for projector generation (see Computational Details). The DFT+U correction on MO projections overcomes limitations of the conventional AO projections and dramatically reduces PBE EDE in $[Mn(CO)_6]^{(3-q)+}$ (Figure 1). By applying DFT+U with MO projectors in increasing values of $U$ that approach the curvature (ca. 6.0–6.3 eV), we largely eliminate the EDE (Figure 1 and Table S2). Reduction in EDE is consistent across the fractional charge line, highlighting the excellent transferability of the MO projectors to the oxidized, $N$-electron (i.e., $q = 0$) $[Mn(CO)_6]^{3+}$ (Figure 1).

In cases with less severe hybridization, the standard DFT+U approach performs reasonably well.[84] As an example, the aqua ligands of the HS $[Fe(H_2O)_6]^{(3-q)+}$ ($q$ = 0 to 1) complex form a weaker ligand field (Figure S1 and Table S3). While there was some evidence of type-i and type-ii errors, including rehybridization upon electron addition in the hexa-aqua complex[84], standard DFT+U successfully eliminates the majority (59%) of PBE EDE at the curvature $U$ value (Figure S1 and Table S4). To consider whether the MO projectors could preserve or even improve upon standard DFT+U for $[Fe(H_2O)_6]^{(3-q)+}$, we construct MO projectors from the reduced state of this complex. Here, the fractional charge line corresponds to quintet $d^6$ Fe(II) to sextet $d^5$ Fe(III), and so we select the minority spin HOMO and four higher, unoccupied MOs from the reduced complex (Table S3 and see Computational Details). The MO projectors naturally incorporate the near-degeneracy of the Fe(3$d$) AOs, with the spin-down fractional electron mostly being added to a single frontier MO, resulting in low type-i errors (Figure S2). Therefore, DFT+U with MO projectors reduces a greater amount of the PBE EDE at the expected (i.e., curvature) $U$ value (Figure S1). While molecular DFT+U improves over the standard approach in both cases, the reduction of EDE at the curvature $U$ value (83%) is somewhat lower for the iron complex than had been achieved by molecular DFT+U for the pathological (i.e., with standard DFT+U) Mn case (94%).

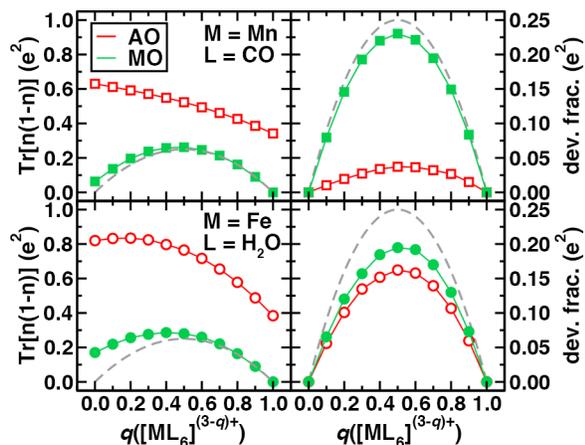

**Figure 2.** (Left) PBE GGA fractionalities (i.e., Tr[**n**(**1-n**)], left) and deviations from the linear admixture (dev. frac., right) from for HS $[Mn(CO)_6]^{(3-q)+}$ (top) and HS $[Fe(H_2O)_6]^{(3-q)+}$ (bottom) for $q$ = 0 to 1. Results are obtained using both AO projectors (red open symbols) and MO projectors from the $q$ = 1 state (green filled symbols). The ideal, atomic limit for all quantities is shown for reference as a gray dashed line.

To better understand the role of projector choice in eliminating EDEs, we analyze the respective PBE fractionality values, Tr[**n**(**1-n**)], with AO or MO projectors as $q$ is varied from 0 to 1, as well as the deviation from a linear admixture of the endpoint values. With standard AO projections in DFT+U for $[Fe(H_2O)_6]^{(3-q)+}$, reduced efficiency of EDE elimination is known to arise from the type-ii error of rehybridization of Fe(3$d$) AOs[84] with fractional electron addition (Figure S3). Both the PBE fractionality and its deviation provide a first-order approximation[84] of the degree to which a DFT+U correction should reduce the EDE (Figure 2). With this analysis, it is evident why the molecular DFT+U improvement is less significant for $[Fe(H_2O)_6]^{(3-q)+}$ than for $[Mn(CO)_6]^{(3-q)+}$ (Figure 2). In the Fe complex, the PBE MO-projector-based fractionality approaching the $q = 0$ limit differs more from the idealized (i.e., zero) value (Figure 2). This results from the fact that the oxidized (i.e., $q = 0$) MOs differ more significantly from the MO projectors obtained on the reduced (i.e., $q = 1$) complex for the iron complex in comparison to the Mn one (Tables S4–S5). The DFT+U correction can be expected to be maximally efficient when deviations of the fractionality approach the idealized limit (i.e., 0.25 $e^2$) at $q = 0.5$.[84] In the AO limit, significant hybridization yields PBE fractionality deviations for both Mn and Fe complexes well below the ideal atomic limit (Figure 2). Although relatively comparable changes in occupations are evident in the AOs of the two complexes, the MO projectors rehybridize less from $q = 1$ to $q = 0$ in the Mn complex, causing the improvements of molecular DFT+U to be somewhat larger (Figure 2). As previously mentioned, the limitation on transferability of the reduced state orbitals due to rehybridization is present for any fixed projector basis and does not provide an advantage to the use of AOs.

In addition to effects on EDE reduction, we should expect the change in projector scheme to alter how DFT+U corrections influence frontier orbital energies. The HOMO and LUMO are intrinsically linked to our definition of EDE. The cubic spline interpolation[26] of the EDE, $E^{dev}(q)$, may be described as a combination of HOMO error (HE) and LUMO error (LE)[84] (i.e., from over- or underestimation of the total-electron result, respectively):

$$E^{dev}(q) = E(q) - \Delta E q = [(\varepsilon_N^{LUMO} - \Delta E)(1-q) + (\Delta E - \varepsilon_{N+1}^{HOMO})q]q(1-q) \qquad (4)$$

where $\Delta E$ is the total energy difference, $E(N+1)-E(N)$, between the $q=1$ and $q=0$ state. The +U correction is incorporated self-consistently with a modification to the potential as:

$$V^U = \sum_{I,nl}\sum_m \frac{U_{nl}^I}{2}(1-2n_{nl,m}^{I\sigma})|\phi_{nl,m}^I\rangle\langle\phi_{nl,m}^I|\,, \qquad (5)$$

When the HOMO and LUMO have the same character as occupied and unoccupied states in the relevant projector scheme, they will be maximally shifted (i.e., by 0.5 eV/eV $U$)[84] down and up in energy, respectively, and EDE will be fully eliminated at the curvature $U$ value. Thus, failure to recover piecewise linearity can also be interpreted as ineffectiveness of DFT+U in correcting HOMO–LUMO gap errors. The lack of effect on HE and LE and thus HOMO–LUMO gaps for standard DFT+U is evident for the $[Mn(CO)_6]^{(3-q)+}$ complex and, to a lesser extent, for $[Fe(H_2O)_6]^{(3-q)+}$ (Figure S4). The $n = \frac{1}{2}$ contribution of the 3$d$ AO in the HOMO and LUMO of $[Mn(CO)_6]^{(3-q)+}$ leads the standard DFT+U potential shift (eq. 5) to have a near-zero effect on the eigenvalues and to even shift the LUMO in the wrong direction (Figure 3 and Table S5). By construction, molecular DFT+U efficiently corrects HOMO and LUMO energies, increasing gaps in both systems (Figure 3 and Figure S4).

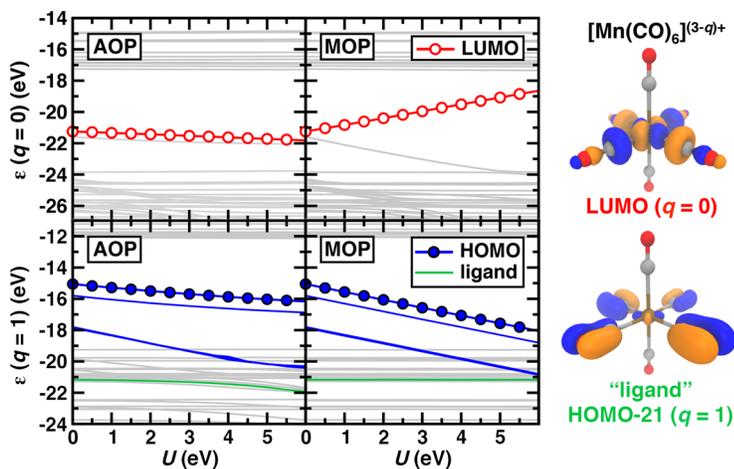

**Figure 3.** MO spin-up eigenvalues (in eV) with applied $U$ (in eV) using DFT+U for HS

$[Mn(CO)_6]^{(3-q)+}$ for $q = 0$ (top) and $q = 1$ (bottom) with atomic orbital projectors, AOP (left) and molecular orbital projectors, MOP (right). The $q = 0$ LUMO eigenvalues (red open circles) and equivalent $q = 1$ HOMO eigenvalues (blue filled circles) are highlighted. Eigenvalues of the other MO states also used to construct real-space MO projectors are indicated with solid blue lines. A ligand-centered, deeper HOMO-21 state is depicted as a solid green line for $q = 1$. Density isosurfaces (|0.002| e/bohr$^3$, blue positive phase, orange negative phase) for the PBE $q = 0$ LUMO and $q = 1$ HOMO-21 states are shown at right.

Additional valence and frontier states are often invoked as catalytic activity descriptors[93-95] or used to interpret electronic properties[96]. Like for the HOMO and LUMO of $[Mn(CO)_6]^{(3-q)+}$, we observe differences in the shifts of other frontier states with the two DFT+U projection approaches (Figure 3). While standard (i.e., AO) DFT+U shifts three valence states (i.e., spin-up HOMO-4 to HOMO-2) with significant atomic character down in energy (ca. -0.4 eV/eV $U$), the HOMO-1 state, like the HOMO, is relatively unaffected (Figure 3). Deeper-lying states with trace atomic 3$d$ character but predominantly consisting of CO π-bonding orbitals are also stabilized with standard DFT+U (Table S6). Molecular DFT+U has a qualitatively different effect on the valence states of $[Mn(CO)_6]^{2+}$ and $[Mn(CO)_6]^{3+}$ (Figure 3). Because the five MO projectors were selected from the $[Mn(CO)_6]^{2+}$ valence states (i.e., majority-spin HOMO-4 to HOMO), all of these eigenvalues in the reduced state are stabilized by the maximum magnitude (i.e., -0.50 eV/eV $U$, Figure 3). Conversely, lower-lying MOs in the reduced complex with residual Mn 3$d$ contributions are orthogonal to these states and unaffected by molecular DFT+U (Figure 3). This observation highlights that, by necessity, some amount of $d$ character is excluded from the MO projectors, causing molecular DFT+U to formally lack rotational invariance in the 3$d$ valence space (Figure 3 and Figure S5 and Tables S7–S8).

The valence states of the oxidized $[Mn(CO)_6]^{3+}$ complex are also stabilized much more significantly with molecular DFT+U than with the standard DFT+U approach due to their good correspondence with the $[Mn(CO)_6]^{2+}$-generated MO projectors (Figure S6). Lower-lying, hybridized states (e.g., HOMO-7) for $[Mn(CO)_6]^{3+}$ are stabilized somewhat more with molecular

DFT+U than with standard AO-based DFT+U but only when these states have comparable character to the five valence states included in the reduced state MO projectors (Table S9 and Figure S7). Overall, distinct shifts occur to valence states with the molecular DFT+U for both reduced and oxidized species.

Consistent with the less pronounced impact of DFT+U projection scheme on EDE reduction for $[Fe(H_2O)_6]^{(3-q)+}$, we observe qualitatively similar frontier orbital energy shifts for the two approaches. For the $[Fe(H_2O)_6]^{2+}$ HOMO and $[Fe(H_2O)_6]^{3+}$ LUMO, eigenvalues shift somewhat more strongly with molecular DFT+U (Figure 4). Since the projectors are obtained from the minority-spin states of $[Fe(H_2O)_6]^{2+}$, the greatest difference between the two schemes can be observed in greater destabilization of unoccupied (i.e., LUMO to LUMO+3) frontier states with molecular DFT+U in comparison to AO projections (Tables S3 and S10–S12). Both standard and molecular DFT+U shift these predominantly (ca. 90%) 3$d$-containing states of the reduced complex above other frontier unoccupied states centered on the ligands or otherwise lacking in 3$d$ character, but molecular DFT+U accomplishes this at lower (ca. 4 eV vs 6–7 eV) $U$ values (Figure 4 and Table S10).

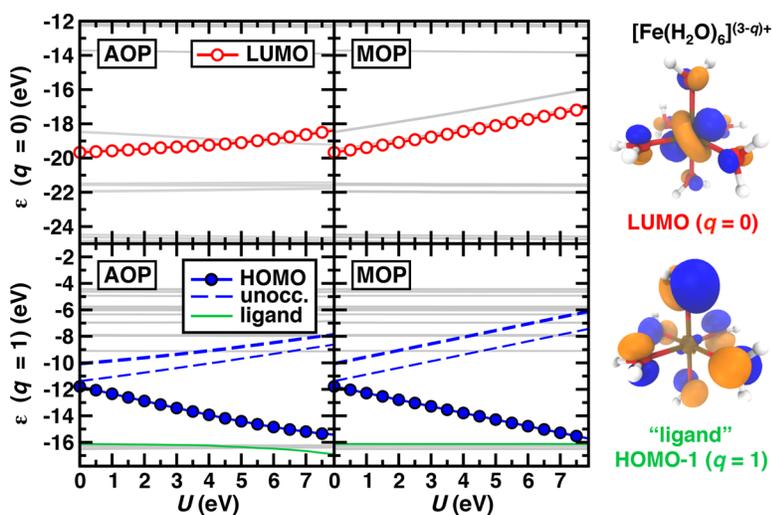

**Figure 4.** MO spin-down eigenvalues (in eV) with applied $U$ (in eV) using DFT+U for HS $[Fe(H_2O)_6]^{(3-q)+}$ for $q = 0$ (top) and $q = 1$ (bottom) with AO (left) and MO (right) projectors. The $q = 0$ LUMO eigenvalues (red open circles) and equivalent $q = 1$ HOMO eigenvalues (blue filled

circles) are highlighted. Eigenvalues of the remaining, unoccupied MO states also used for constructing real-space MO projectors are indicated with dashed blue lines. A ligand-centered, HOMO-1 state is depicted as a solid green line for $q = 1$. Density isosurfaces ($|0.002|$ e/bohr$^3$, blue positive phase, orange negative phase) for the PBE $q = 0$ LUMO and $q = 1$ HOMO-1 states are shown at right.

For the oxidized $[Fe(H_2O)_6]^{3+}$ complex, low-lying unoccupied minority-spin states including the LUMO (i.e., LUMO to LUMO+4) are predominantly (ca. 80–90%) 3$d$-centered with some ligand participation (Table S13). However, the near-degeneracy of the spin-down orbitals leads to partial occupation of 3$d$ e$_g$ atomic states (i.e., $d_{x^2-y^2}$ and $d_{z^2}$), causing standard DFT+U to have either a weakly destabilizing effect (i.e., LUMO through LUMO+2) or a counterintuitive stabilizing effect (i.e., for LUMO+3 and LUMO+4) on some states (Table S13). Molecular DFT+U instead shifts upwards all of the 3$d$-containing unoccupied states in a more consistent manner, although it does so at slightly lower efficiency than for the reduced complex from which the molecular projectors were extracted (Figure 4 and Table S13).

Finally, to validate the general ability of MO projectors to improve upon pathological cases while preserving EDE reduction in cases where standard DFT+U performed well, we generated MO projectors and expanded our analysis to a diverse range of eight additional transition-metal complexes (Figure 5 and Tables S14–S21). As a general strategy for MO projector generation, we use the reduced state of the complex and select its HOMO and four additional states around the HOMO based on the $d$ electron count of the isolated ion (e.g., HOMO to LUMO+3 for $d^1$ Sc(II), see Computational Details and Text S1).

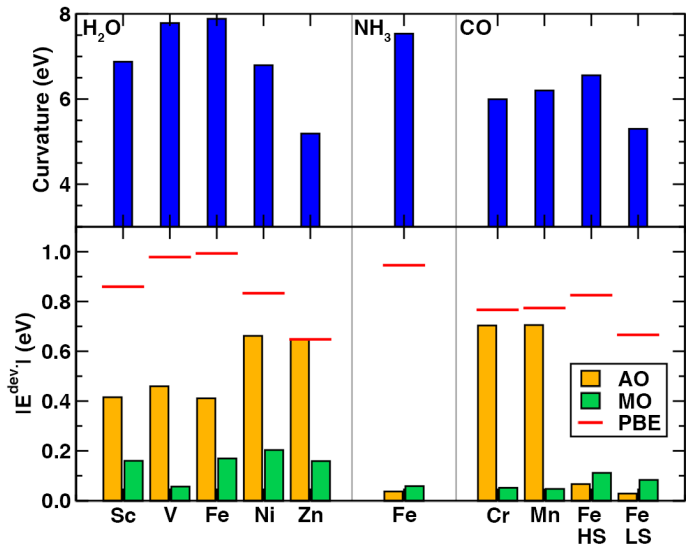

**Figure 5.** Homoleptic octahedral complexes grouped by ligand (indicated top inset) and metal center (indicated on axis) for $[M(L)_6]^{(3-q)+}$, where all complexes are in their HS state except for Fe(CO)$_6$ which is studied in both LS and HS states. Curvature values (in eV) from PBE are shown at top, and the unsigned deviation from linearity ($|E^{dev.}|$, in eV) at $q = 0.5$ is shown for PBE (red horizontal line) and DFT+U with AO (orange bars) or MO (green bars) projectors. The DFT+U results are obtained at the $U$ value corresponding to the curvature.

Regardless of $d$ filling and the magnitude of initial EDE (i.e., the PBE curvature), molecular DFT+U consistently improves upon the standard DFT+U approach for all hexa-aqua complexes studied (Figure 5 and Figures S8–S12 and Table S22). For the cases where standard DFT+U performs relatively well (i.e., Sc, V, or Fe), the relevant HOMO and LUMO states have strong (ca. 70–90%) atomic $d$ character (Tables S4, S23–S24). Conversely, rehybridization in Ni with electron addition as well as lower $d$ character (ca. 33%) in the LUMO means the standard DFT+U approach has a limited effect in reducing EDE for this complex (Table S25 and Figure S13). As an extreme example[89], the filling of the $d$ shell in the $[Zn(H_2O)_6]^{(3-q)+}$ complex ($q = 0$ to 1) leads to fully occupied AOs and no 3$d$ contribution in the HOMO or LUMO (Table S21 and Figures S12 and S14). The result is that the standard DFT+U has no effect on the PBE EDE for Zn, whereas only a small degree of EDE remains after correction with molecular DFT+U, comparable to that for other hexa-aqua complexes (Figure 5 and Figure S12). Molecular DFT+U therefore represents a general approach to EDE reduction in these systems, with the caveat that rehybridization may limit EDE reduction at the expected (i.e., curvature) $U$ value given the need to choose a single projector reference state.

For hexa-carbonyl complexes, Cr and Mn are both pathological cases with standard DFT+U, whereas standard DFT+U nearly completely eliminates EDE for Fe(CO)$_6$ in both the HS and LS states (Figure 5). This difference can be traced to the fact that for Cr or Mn the frontier orbitals are strongly hybridized between metal and ligand, whereas for Fe they are centered on the metal (Tables S5 and S26–S28). For all hexa-carbonyl complexes studied, molecular DFT+U again improves the pathological cases and preserves EDE reduction in cases where standard DFT+U already performed well (Figures 1 and 5 and Figures S15–S17). For the intermediate ligand field strength hexa-ammine iron complexes, the frontier states are primarily metal-centered, and good EDE reduction is achieved with either projector scheme (Figure S18 and Table S29). Comparing four iron complexes, the PBE EDE (i.e., curvature) decreases with increasing field strength and for the LS complex in comparison to its HS counterpart (Figure 5). However, the initial curvature or spin state choice has only an indirect role in the residual EDE after application of either DFT+U scheme, with molecular DFT+U fairly consistently reducing EDE (Figure 5). Thus, this molecular approach to DFT+U generalizes to chemistry where standard DFT+U would be ineffective at correcting delocalization error.

In summary, we have demonstrated how the widely applied DFT+U correction can fail to reduce the delocalization error of the underlying semi-local functional in molecules and materials characterized by covalent bonding. Because correct descriptions of the electronic structure of many transition-metal complexes and correlated materials are expected to involve significant delocalized bonding after recovery of exact conditions (e.g., piecewise linearity), a distinct approach is required. By preserving the simple quadratic functional form of DFT+U but adopting an alternative valence state MO basis for calculating occupations in DFT+U, we consistently improve over the standard DFT+U approach in the elimination of delocalization error for cases previously identified to be pathological. While a formal loss of rotational invariance had no negative impact in the systems studied, further study is warranted when extending the molecular DFT+U approach to catalytic bond rearrangement. This approach does not worsen the cases where the standard DFT+U formulation was already successful. Importantly, these observations are general across a range of ligand field strengths, $d$-filling (i.e., including closed-shell $d^{10}$ Zn), and spin states. Since the MO projectors can be generated automatically, at relatively low cost, and are moderately transferable across oxidation states, we expect our approach will be of utility in improving the quality of the large number of high-throughput screening workflows that already employ standard DFT+U.

**Computational Details.** Spin-polarized DFT calculations were carried out with the plane-wave periodic boundary condition code Quantum-ESPRESSO v5.1[97] using the Perdew–Burke–Ernzerhof (PBE) semilocal GGA[11] xc functional. Ultrasoft pseudopotentials[98-99] (USPPs) obtained from the Quantum-ESPRESSO website[100] were employed throughout, with semi-core (i.e., 3s or 3p) states included in the valence for early and mid-row (i.e., Sc–Fe) metals (Table S30). We employed plane-wave cutoffs of 30 Ry for the wavefunction and 300 Ry for the charge density, as in prior work[84]. All complexes were placed in a 14.8 Å cubic box to ensure sufficient vacuum, and the Martyna–Tuckerman scheme[101] was employed to eliminate periodic image effects and enable comparison of total energies at varying charge. More than 15 and up to a maximum of 26 unoccupied states (i.e., bands) were included for all calculations. To improve self-consistent field (SCF) convergence, the mixing factor was reduced to 0.4 from its default value, and the convergence threshold for the SCF energy error was loosened to $9\times10^{-6}$ Ry.

Single-point calculations were carried out on the optimized geometries of homoleptic octahedral transition-metal complexes from Ref. [84] that had a net charge of +3, with geometries adjusted in this work so that the metal was at the center of the box (see Supporting Information for geometries). Fractional charge calculations were carried out for net charges from +3 to +2 in increments of 0.1 e, with manually adjusted band occupations using the "from_input" command in Quantum-ESPRESSO. Except for the low-spin $Fe(CO)_6$, all calculations employed the high-spin ground state of the isolated metal atom, as obtained from the National Institute of Standards and Technology (NIST) database[102] for the M(II) and M(III) centers (Table S31).

Plane-wave eigenstates were transformed to their corresponding real-space Wannier function[103] localized molecular orbitals using the pmw.x utility in Quantum-ESPRESSO. Five states were selected in order on the reduced state of the complex. To choose the starting index (i.e., using the "first_band" keyword), we noted the band for which the electron would be added or removed as well as four additional states based on the electron configuration. For reduced complexes with less than or equal to half-filled *d*-shells (i.e., $d^x$, $x$ = 1 to 5), majority spin MOs starting from the HOMO-($x$-1) and four higher energy states were used. For reduced complexes with later transition metals (i.e., $d^x$, $x$ = 6 to 10), minority spin MOs corresponding to the HOMO-($x$-6) and four higher energy states were used. The precise band numbers depended on whether semi-core states were included in the pseudopotentials (Tables S30–S31). Density isosurfaces of molecular orbitals were plotted at ±0.002 e/bohr$^3$ using the pp.x utility of Quantum-ESPRESSO and visualized with VMD[104].

## ASSOCIATED CONTENT

**Supporting Information**. The following files are available free of charge.

Overview of approach for construction of the real-space MO projectors; electronic configurations with term-symbols for all metal centers; AO contributions of the real-space MO projectors for all complexes; total $3d$ AO contributions for the MO projectors and for all the low-lying states of $[Mn(CO)_6]^{2+}$ and $[Fe(H_2O)_6]^{2+}$; PBE-level frontier eigenvalues and curvature $U$ estimates; AO contributions to the HOMO ($q = 1$) and LUMO ($q = 0$) states; variation of PBE-level $3d$ AO and projector MO occupations upon fractional charge addition for $Mn(CO)_6$, $Fe(H_2O)_6$, $Ni(H_2O)_6$ and $Zn(H_2O)_6$; deviations from linearity with $+U$ using AO and MO projectors for complexes other than $Mn(CO)_6$; HOMO error and LUMO error variation with $+U$ using AO and MO projectors for $Mn(CO)_6$ and $Fe(H_2O)_6$; variation of AO contributions to a low-lying state of $[Mn(CO)_6]^{2+}$ with $+U$ using AO projectors; PBE-level AO contributions, density isosurface plots and eigenvalue shifts with $+U$ using AO and MO projectors for a non-projector high-energy state of $[Mn(CO)_6]^{2+}$; density isosurface plots of comparable frontier and low-lying states of $[Mn(CO)_6]^{3+/2+}$; PBE-level AO contributions and eigenvalue shifts with $+U$ using AO and MO projectors for low-lying states of $[Mn(CO)_6]^{3+}$; frontier eigenvalue shifts with $+U$ using AO and MO projectors for $[Fe(H_2O)_6]^{2+}$; PBE-level AO contributions and eigenvalue shifts with $+U$ using AO and MO projectors for a non-projector high-energy state of $[Fe(H_2O)_6]^{2+}$; PBE-level AO contributions and eigenvalue shifts with $+U$ using AO and MO projectors for frontier states of $[Fe(H_2O)_6]^{3+}$; interpolation of EDE for $Sc(H_2O)_6$ and $V(H_2O)_6$; comparison of reductions in EDE between AO and MO projectors for all aqua complexes; pseudopotentials used for all calculations (PDF)

Geometries (.xyz files) with metal-centers repositioned at the center of a periodic box and example input files (ZIP)

AUTHOR INFORMATION

**Notes**

The authors declare no competing financial interests.

ACKNOWLEDGMENT

The authors acknowledge primary support by the Department of Energy under grant number DE-SC0018096. Additional support was provided by the National Science Foundation grants CBET-1704266 and CBET-1846426 and by the Office of Naval Research under grant number N00014-18-1-2434. H.J.K. holds a Career Award at the Scientific Interface from the Burroughs Wellcome Fund and an AAAS Marion Milligan Mason Award, which supported this work. The authors thank Adam H. Steeves for providing a critical reading of the manuscript.